\documentclass[12pt]{article}
\usepackage{graphicx}
\usepackage{amsmath}
\usepackage{amssymb}
\usepackage{caption2}
\begin{document}
\bibliographystyle{prsty}
\begin{center}
{\large {\bf \sc{  Analysis of the $X(1835)$ as a baryonium state with Bethe-Salpeter equation }}} \\[2mm]
Zhi-Gang Wang \footnote{E-mail,wangzgyiti@yahoo.com.cn.  }    \\
$^{1}$ Department of Physics, North China Electric Power University,
Baoding 071003, P. R. China
\end{center}

\begin{abstract}
In this article, we take  the $X(1835)$ as a pseudoscalar baryonium
state,  and calculate the mass spectrum of the baryon-antibaryon
bound states $p\bar{p}$, $\Sigma\bar{\Sigma}$, $\Xi\bar{\Xi}$, and
$\Lambda\bar{\Lambda}$ in the framework of the Bethe-Salpeter
equation with a phenomenological potential. The numerical results
indicate  the $p\bar{p}$, $\Sigma\bar{\Sigma}$ and $\Xi\bar{\Xi}$
bound states  maybe exist, and the $X(1835)$ can be tentatively
identified as the $p\bar{p}$ bound state.
\end{abstract}



\section{Introduction}
In 2003, the BES collaboration  observed a significant narrow
near-threshold enhancement in the proton-antiproton ($p\bar{p}$)
invariant mass spectrum  in the radiative decay $J/\psi\to\gamma
p\overline{p}$ \cite{BES03}.
 The enhancement can be fitted  with either an $S$- or $P$-wave Breit-Wigner
resonance function. In the case of the $S$-wave fitted form, the
 mass and the width are $M = 1859 {}^{+3}_{-10}
{}^{+5}_{-25} \,\rm{MeV}$ and $\Gamma < 30 \,\rm{MeV}$ respectively.
In 2005, the BES collaboration observed a resonance  state $X(1835)$
in the $\eta^{\prime}\pi^+\pi^-$ invariant mass spectrum in the
process $J/\psi\to\gamma\pi^{+}\pi^{-}\eta^{\prime}$  with the
Breit-Wigner mass $M=(1833.7\pm 6.2\pm 2.7)\,\rm{MeV}$  and  the
width $\Gamma=(67.7\pm 20.3\pm 7.7)\,\rm{MeV}$ respectively
\cite{BES05}. Many theoretical works were stimulated to interpret
the nature and the structure of the new particle, such as the
$p\bar{p}$ bound state
\cite{pp-Rosner,pp-Datta,pp-Zou,pp-Liu,pp-Chang,pp-Sibirtsev,pp-Yan,pp-Yan-2,pp-Zhu,pp-Wang,pp-Ma,pp-Dedonder,pp-Ma-2},
the pseudoscalar glueball \cite{gb-Kochelev,gb-Li,gb-Hao,gb-He}, and
the radial excitation of the $\eta'$
\cite{eta-Huang,eta-Klempt,eta-Li}, etc.

In this article, we take  the $X(1835)$  as a baryonium with the
quantum numbers $J^{PC}=0^{-+}$, and calculate the mass spectrum of
the baryon-antibaryon bound states $p\bar{p}$, $\Sigma\bar{\Sigma}$,
$\Xi\bar{\Xi}$, and $\Lambda\bar{\Lambda}$ with the Bethe-Salpeter
equation \cite{BS51,BS69}.
 The Bethe-Salpeter equation  with phenomenological potentials is a powerful theoretical tool
in studying bound states and has given many successful descriptions
of the hadron  properties \cite{BS69,Roberts94,Roberts00}.

The article is arranged as follows:  we solve the Bethe-Salpeter
equation  for the baryon-antibaryon bound  states  in Sec.2; in
Sec.3, we present the numerical results and discussions; and Sec.4
is reserved for our conclusions.

\section{Bethe-Salpeter equation}
The Bethe-Salpeter equation  is a conventional approach in dealing
with the two-body relativistic bound state problems
\cite{BS51,BS69,Roberts94,Roberts00}. We write down the ladder
Bethe-Salpeter equation for  the pseudoscalar $p\bar{p}$ bound
state,
\begin{eqnarray}
 S^{-1}\left(q+\frac{
P}{2}\right)\chi(q,P)S^{-1}\left(q-\frac{ P}{2}\right)&=&\int
\frac{d^4
k}{(2\pi)^4}\gamma_5 \chi(k,P) \gamma_5 G(q-k)\, ,\\
S^{-1}\left(q\pm \frac{P}{2}\right)&=&i\left(\gamma \cdot q\pm
\frac{\gamma \cdot P}{2}\right)+M_p \, , \nonumber
\end{eqnarray}
where the $P_\mu$ is the four-momentum of the center of mass of the
$p\bar{p}$ bound state, the $q_\mu$ is the relative four-momentum
between the proton and antiproton,
 $\gamma_{5}$ is the bare baryon-meson vertex,
the $\chi(q,P)$ is the Bethe-Salpeter amplitude of the $p\bar{p}$
bound state, and the $G(q-k)$ is the interaction kernel. With a
simple replacement of the corresponding parameters, we can obtain
the Bethe-Salpeter equations
 for the $\Sigma\bar{\Sigma}$, $\Xi\bar{\Xi}$ and
$\Lambda\bar{\Lambda}$ bound states if they exist.

In the flavor $SU(3)$ symmetry limit,  the interactions among the
octet baryons and the pseudoscalar mesons can be described by the
lagrangian $\cal L$,
\begin{eqnarray}
{\cal L} = \sqrt2 \left( D {\rm Tr} \left(\bar B
\left\{P,B\right\}_{+}\right)+ F {\rm Tr} \left(\bar B \left[ P,B
\right]_{-} \right) \right)\, ,
\end{eqnarray}
where
\begin{eqnarray}
B &=& \left(
\begin{array}{ccc}
    \frac{1}{\sqrt{2}} \Sigma^0 + \frac{1}{\sqrt{6}} \Lambda & \Sigma^+ & p \\
    \Sigma^- & - \frac{1}{\sqrt{2}}\Sigma^0 + \frac{1}{\sqrt{6}} \Lambda & n \\
    \Xi^- & \Xi^0 & - \frac{2}{\sqrt{6}} \Lambda
\end{array} \right) \, ,\nonumber\\
P &=& \left(
\begin{array}{ccc}
    \frac{1}{\sqrt2} \pi^0 + \frac{1}{\sqrt{6}}\eta & \pi^+ & K^+ \\
    \pi^- & - \frac{1}{\sqrt2} \pi^0 + \frac{1}{\sqrt6} \eta & K^0 \\
    K^- & \bar K^0  &- \frac{2}{\sqrt6} \eta
\end{array}
\right) \, ,
\end{eqnarray}
and the $D$ and $F$ are two parameters for the coupling constants.
From the lagrangian, we can obtain
\begin{eqnarray}
g_{\pi^0 p p} &=& -g_{\pi^0 n n}=D + F\, , \,\, g_{\pi^0 \Sigma^+
\Sigma^+} =- g_{\pi^0 \Sigma^- \Sigma^-} =2 F\, ,
\nonumber \\
g_{\pi^0 \Xi^- \Xi^-} &=& -g_{\pi^0 \Xi^0 \Xi^0}=D-F\, , \,\,
g_{\eta p p} = g_{\eta nn}=-\frac{D-3F}{\sqrt3} \, ,
\nonumber \\
g_{\eta \Sigma^+ \Sigma^+} &=& g_{\eta \Sigma^- \Sigma^-} =g_{\eta
\Sigma^0 \Sigma^0} =- g_{\eta \Lambda \Lambda}=\frac{2D}{\sqrt3}
\, , \nonumber \\
g_{\eta \Xi^-\Xi^-} &=& g_{\eta \Xi^0\Xi^0}=-\frac{D+3F}{\sqrt3}\, ,
\end{eqnarray}
and write down the kernel $G(q-k)$ explicitly,
\begin{eqnarray}
G(q-k)&=&\frac{g^2(q-k)C_\pi}{(q-k)^2+m_\pi^2}+\frac{g^2(q-k)C_\eta}{(q-k)^2+m_\eta^2}
\, ,
\end{eqnarray}
where the coefficients  $C_\pi=(1+\alpha)^2$, $4\alpha^2$,
$(1-\alpha)^2$, $0$ and $C_\eta=\frac{(1-3\alpha)^2}{3}$,
$\frac{4}{3}$, $\frac{(1+3\alpha)^2}{3}$, $\frac{4}{3}$ for the
$p\bar{p}$, $\Sigma\bar{\Sigma}$, $\Xi\bar{\Xi}$,
$\Lambda\bar{\Lambda}$ bound states respectively; $g^2(k)=D^2$ and
$\alpha=\frac{F}{D}$. In this article, we choose the value
$\alpha=0.6$ from  the analysis of the hyperon semi-leptonic decays
\cite{FD}, and take the coupling constant $g^2(k)$ as a modified
Gaussian distribution, $g^2(k)=A
\left(\frac{k^2}{\mu^2}\right)^2\exp\left(-
\frac{k^2}{\mu^2}\right)$, where the strength  $A$ and the
distribution width $\mu^2$ are two free parameters. The ultraviolet
behavior of the modified Gaussian distribution warrants the integral
in the Bethe-Salpeter equation is  convergent.

We can perform the Wick rotation analytically and continue  $q$, $k$
into the Euclidean region.
 The Euclidean Bethe-Salpeter amplitude of the pseudoscalar $p\bar{p}$ bound state can be decomposed as
 \begin{eqnarray}
 \chi(q,P)&=&
 \gamma_5\left\{F(q,P)+i\!\not\!{P}F_1(q,P)
 +i\!\not\!{q}F_2(q,P)+\left[\!\not\!{q},\!\not\!{P}\right]F_3(q,P)
 \right\}
 \, ,
 \end{eqnarray}
 due to  Lorentz covariance \cite{Roberts94}. In the coordinate space, the
 Bethe-Salpeter amplitude is defined by
 $\chi_{\alpha\beta}(x,y)=\langle0|T\left[ q_\alpha(y)
 \bar{q}_\beta(x)\right]|X\rangle$, where the $q(x)$ is the interpolating  current of the proton,
  $q(x)=\epsilon^{ijk}  u^T_i(x)C\gamma_\mu u_j(x) \gamma_5 \gamma^\mu d_k(x)$.
We can perform the Fierz re-ordering in the Dirac spinor space to
obtain the following identity,
\begin{eqnarray}
q_\alpha(y) \bar{q}_\beta(x)&=&-\frac{1}{4}
\delta_{\alpha\beta}\bar{q}(x)q(y)
-\frac{1}{4}(\gamma^\mu)_{\alpha\beta}\bar{q}(x)\gamma_\mu
q(0) -\frac{1}{8}(\sigma^{\mu\nu})_{\alpha\beta}\bar{q}(x)\sigma_{\mu\nu}q(y) \nonumber\\
&&+\frac{1}{4}(\gamma^\mu
\gamma_5)_{\alpha\beta}\bar{q}(x)\gamma_\mu \gamma_5
q(y)+\frac{1}{4} (i \gamma_5)_{\alpha\beta}\bar{q}(x)i \gamma_5 q(y)
\, .
\end{eqnarray}
 The QCD sum rules indicate  that the couplings of the axial-vector currents and
the pseudoscalar currents to the octet pseudoscalar mesons are much
stronger than other interpolating quark currents \cite{NarisonBook}.
In this article, we can take the approximation
\begin{eqnarray}
 \chi(q,P)&=&
 \gamma_5\left\{F(q,P)+i\!\not\!{P}F_1(q,P)\right\} \, ,
 \end{eqnarray}
 for simplicity.

 Multiplying both sides of the Bethe-Salpeter equation   by
 $\gamma_5\left[\!\not\!{q},\!\not\!{P}\right]$ and doing the trace
 in the Dirac spinor space, we can obtain an  simple relation
 $F=2M_pF_1$, the amplitudes $F(q,P)$ and $F_1(q,P)$ are not independent.

 The Bethe-Salpeter  amplitude can be written  as
\begin{eqnarray}
 \chi(q,P)&=&
 \gamma_5\left(1+\frac{i\!\not\!{P}}{2M_p}\right)F(q,P)  \, ,
 \end{eqnarray}
 and the Bethe-Salpeter equation can be projected into the following form,
 \begin{eqnarray}
 \left( q^2+M_p^2+\frac{P^2}{4}\right)F(q,P)&=&\int \frac{d^4k}{(2\pi)^4}
  F(k,P)G(q-k) \, .
 \end{eqnarray}
We can introduce a parameter $\lambda(P^2)$ and solve  above
equation as an eigenvalue problem.  If there really exists  a bound
state in the pseudoscalar channel, the mass of the $X(1835)$ can be
determined by the condition $\lambda(P^2=-M_{X}^2)=1$,
\begin{eqnarray}
\left( q^2+M_p^2+\frac{P^2}{4}\right)F(q,P)&=&\lambda(P^2)\int
\frac{d^4k}{(2\pi)^4}
  F(k,P)G(q-k) \, .
\end{eqnarray}
In the limit $q^2=0$, we can obtain an simple relation  for the
$p\bar{p}$ bound state,
\begin{eqnarray}
M_X^2<4M_p^2 \, ,
\end{eqnarray}
i.e. the bound energy should be negative, $E_X=2M_p-M_X<0$. The
Bethe-Salpeter equations for other bound states are treated with the
same routine.

\section{Numerical results and discussions}
The input parameters are taken as $m_\pi=135\,\rm{MeV}$,
$m_\eta=548\,\rm{MeV}$, $M_p=938.3\,\rm{MeV}$,
$M_{\Sigma^+}=1189.4\,\rm{MeV}$, $M_{\Xi^-}=1321.7\,\rm{MeV}$,
$M_{\Lambda}=1115.7\,\rm{MeV}$, and $M_{X(1835)}=1833.7\,\rm{MeV}$
from the Particle Data Group  \cite{PDG}. The strength  $A$ and the
distribution width $\mu^2$ are
 free parameters, we take the values $A=215$ and
$\mu=200\,\rm{MeV}$ for the $p\bar{p}$ bound state. Furthermore, we
take the simple replacements $\mu\rightarrow \mu
\frac{M^2_{\Sigma}}{M_p^2}$, $ \mu \frac{M^2_{\Xi}}{M_p^2}$ and $
\mu \frac{M^2_{\Lambda}}{M_p^2}$ to take into account the flavor
$SU(3)$ breaking effects for the $\Sigma\bar{\Sigma}$,
$\Xi\bar{\Xi}$ and $\Lambda\bar{\Lambda}$ bound states respectively.

We solve the Bethe-Salpeter equations  as  an eigen-problem
numerically by direct iterations, and observe the convergent
behaviors  are  very good. For the $p\bar{p}$, $\Sigma\bar{\Sigma}$
and $\Xi\bar{\Xi}$ bound states, there
 exists  a solution with $\lambda(P^2=-M_{X}^2)=1$ and $E_X<0$.
 On the other hand, we cannot obtain a solution to satisfy the condition
  $\lambda(P^2=-M_{X}^2)=1$ for the $\Lambda\bar{\Lambda}$ bound state,
  and have to  resort to the fine-tune mechanism
    by  introducing the coupling $g_{\eta'\Lambda\Lambda}$
  between the $\eta'$ meson and the
   $\Lambda$ baryon. For example, we can obtain a $\Lambda\bar{\Lambda}$ bound state
   with the bound energy  $E=-30\,\rm{MeV}$ with the value
   $g_{\eta'\Lambda\Lambda}^2=5.3g_{\eta\Lambda\Lambda}^2$, such as
   a fine-tune solution should not be taken seriously.
    The numerical results for the Bethe-Salpeter amplitudes
are shown in Fig.1 and the values of the bound states are presented
in Table 1.

\begin{table}
\begin{center}
\begin{tabular}{|c|c|c|c|c|}
\hline\hline
 &$p\bar{p}$&$\Sigma\bar{\Sigma}$&$\Xi\bar{\Xi}$&$\Lambda\bar{\Lambda}$\\ \hline
 $M_X[\rm{MeV}]$& $1833.7$ & $2317.8$ & $2612.4$ &  $2201.4$ \\ \hline
 $E_X[\rm{MeV}]$& $-42.9$& $-61.0$ &  $-31.0$& $-30$ \\ \hline
 \hline
\end{tabular}
\end{center}
\caption{ The masses $M_X$ and the bound energies $E_X$   for the
baryon-antibaryon  bound states.}
\end{table}

\begin{figure}
 \centering
 \includegraphics[totalheight=7cm,width=8cm]{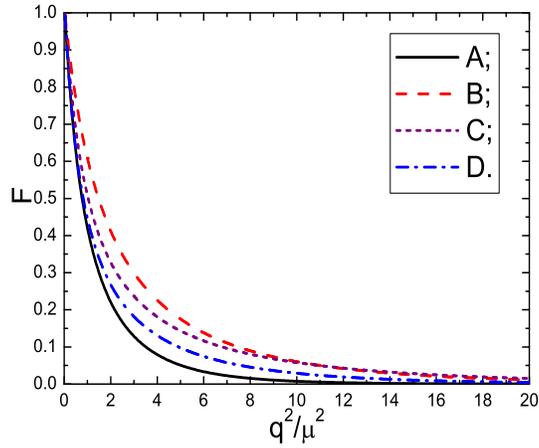}
  \caption{The Bethe-Salpeter amplitudes of the bound states, $A$, $B$, $C$ and $D$ denote the $p\bar{p}$,
  $\Sigma\bar{\Sigma}$, $\Xi\bar{\Xi}$ and
$\Lambda\bar{\Lambda}$ respectively.  }
\end{figure}

If those bound states  exist indeed, they can be produced in the
radiative $J/\psi$ decays, i.e. $J/\psi\rightarrow\gamma gg$,
 $gg+q\bar{q}\rightarrow
p\bar{p},\,\Sigma\bar{\Sigma},\,\Xi\bar{\Xi},\,\Lambda\bar{\Lambda}$,
those bound states can decay to the $\eta\pi\pi$, $\eta K\bar{K}$,
$\eta'\pi\pi$, $\eta' K\bar{K}$, $\eta'\eta\eta$, $\eta'\eta'\eta$,
$\eta\eta\eta$ final states. We can search for  those bound states
in the $\eta\pi\pi$, $\eta K\bar{K}$, $\eta'\pi\pi$, $\eta'
K\bar{K}$, $\eta'\eta\eta$, $\eta'\eta'\eta$, $\eta\eta\eta$
invariant mass distributions   in the radiative decays  of the
$J/\psi$ at the BESIII \cite{BESIII}.

\section{Conclusion}
In this article, we take  the $X(1835)$ as a pseudoscalar baryonium
state and calculate the mass spectrum of the baryon-antibaryon bound
states
$p\bar{p},\,\Sigma\bar{\Sigma},\,\Xi\bar{\Xi},\,\Lambda\bar{\Lambda}$
in the framework of the Bethe-Salpeter equation with a
phenomenological potential. The numerical results indicate that
 the $p\bar{p},\,\Sigma\bar{\Sigma},\,\Xi\bar{\Xi}$ bound states maybe exist, and
 the $X(1835)$ can be tentatively identified as the $p\bar{p}$
 bound state. The other  bound  states maybe observed in
 the future  at the BESIII.

\section*{Acknowledgments}
This  work is supported by National Natural Science Foundation,
Grant Number 10775051, and Program for New Century Excellent Talents
in University, Grant Number NCET-07-0282, and the Fundamental
Research Funds for the Central Universities.


\begin{thebibliography}{99}
\bibitem{BES03} J. Z. Bai et al.,   Phys. Rev. Lett. {\bf 91} (2003) 022001 .
\bibitem{BES05} M. Ablikim et al., Phys. Rev. Lett. {\bf 95} (2005) 262001.

\bibitem{pp-Rosner} J. L. Rosner,  AIP Conf. Proc. {\bf 815} (2006) 218.

\bibitem{pp-Datta} A. Datta and P. J. O'Donnell, Phys. Lett. {\bf B567} (2003) 273.

\bibitem{pp-Zou} B. S. Zou and H. C. Chiang, Phys. Rev. {\bf D69} (2004) 034004.

\bibitem{pp-Liu} X. Liu, X. Q. Zeng, Y. B. Ding, X. Q. Li, H. Shen and P. N. Shen, hep-ph/0406118.

\bibitem{pp-Chang} C. H. Chang and H. R. Pang,  Commun. Theor. Phys. {\bf 43} (2005) 275.

\bibitem{pp-Sibirtsev} A. Sibirtsev, J. Haidenbauer, S. Krewald, Ulf-G. Meissner and A. W. Thomas, Phys. Rev. {\bf D71} (2005) 054010.

\bibitem{pp-Yan} M. L. Yan, S. Li, B. Wu and B. Q. Ma,  Phys. Rev. {\bf D72} (2005) 034027.

\bibitem{pp-Yan-2} G. J. Ding and M. L. Yan, Phys. Rev. {\bf C72} (2005) 015208.

\bibitem{pp-Zhu}  S. L. Zhu  and C. S. Gao, Commun. Theor. Phys. {\bf 46} (2006) 291.

\bibitem{pp-Wang} Z. G. Wang and S. L. Wan, J. Phys. {\bf G34} (2007) 505.

\bibitem{pp-Ma} Y. L. Ma, J. Phys. {\bf G36} (2009) 055004.

\bibitem{pp-Dedonder} J. P. Dedonder, B. Loiseau,  B. El-Bennich and  S. Wycech, Phys. Rev. {\bf  C80} (2009) 045207.

\bibitem{pp-Ma-2} G. Y. Chen,   H. R. Dong and   J. P. Ma,   Phys. Rev. {\bf D78} (2008) 054022.

\bibitem{gb-Kochelev} N. Kochelev and D. P. Min, Phys. Rev. {\bf D72} (2005) 097502.

\bibitem{gb-Li} B. A. Li, Phys. Rev. {\bf D74} (2006) 034019.

\bibitem{gb-Hao} G. Hao, C. F. Qiao and A. L. Zhang, Phys. Lett. {\bf B642} (2006) 53.

\bibitem{gb-He}  X. G. He, X. Q. Li, X. Liu and J. P. Ma,  Eur. Phys. J. {\bf C49} (2007) 731.


\bibitem{eta-Huang} T. Huang and S. L. Zhu, Phys. Rev. {\bf D73} (2006) 014023.

\bibitem{eta-Klempt}  E. Klempt and A. Zaitsev, Phys. Rept. {\bf 454} (2007) 1.

\bibitem{eta-Li} D. M. Li and B. Ma, Phys. Rev. {\bf D77} (2008) 074004.

\bibitem{BS51} E. E. Salpeter and H. A. Bethe, Phys. Rev. {\bf 84} (1951) 1232.

\bibitem{BS69} N. Nakanishi, Suppl. Prog. Theor. Phys. {\bf 43} (1969) 1.

\bibitem{Roberts94} C. D. Roberts and A. G. Williams, Prog. Part. Nucl. Phys. {\bf 33} (1994)  477.

\bibitem{Roberts00} C. D. Roberts and S. M. Schmidt, Prog. Part. Nucl. Phys. {\bf 45} (2000) S1.

\bibitem{FD} P. G. Ratcliffe, Phys. Lett. {\bf B365} (1996) 383.

\bibitem{NarisonBook} S. Narison, Camb. Monogr. Part. Phys. Nucl. Phys. Cosmol. {\bf 17} (2002) 1.

\bibitem{PDG} C. Amsler et al.,  Phys. Lett. {\bf B667} (2008) 1.

\bibitem{BESIII} D. M. Asner et al., Int. J. Mod. Phys. {\bf A24} (2009)  Supp 1.

\end{thebibliography}
\end{document}